# Atomic size effect in impurity indued grain boundary embrittlement

W. T. Geng*, A. J. Freeman†, G. B. Olson‡

\* Department of Physics, Qingdao University, Qingdao 266071, China

geng@qdu.edu.cn

† Department of Physics & Astronomy, Northwestern University, Evanston, Illinois 60208, USA

‡ Department of Materials Science & Engineering, Northwestern University, Evanston, Illinois 60208, USA

Bismuth segregated to the grain boundary in Cu is known to promote brittle fracture of this material. Schweinfest *et al.*[1] reported first-principles quantum mechanical calculations on the electronic and structural properties of a Cu grain boundary with and without segregated Bi and argue that the grain boundary weakening induced by Bi is a simple atomic size effect. But their conclusion is invalid for both Bi and Pb because it fails to distinguish the chemical and mechanical (atomic size) contributions, as obtained with our recently developed first-principles based phenomenological theory.[2]

Significantly, Schweinfest *et al.*'s work rules out embrittlement mechanisms *bond stiffing* at the grain boundary core and *charge transfer* from Cu-Cu bonds to the Bi-Cu bonds. With the observation of volume expansion and Cu-$d$ band narrowing at the grain boundary, they attribute the decrease of the work needed to cleave the Cu $\Sigma$19a grain boundary fully to the large atomic size of Bi. This conclusion, however, is based on an incomplete approach. According to Rice-Wang theory,[3] the potency of a segregation impurity in reducing the work needed to cleave a brittle boundary is a linear function of the difference in *binding energies* for that impurity at the grain boundary and the free surface. Thus, without *bond stiffing* and *charge transfer*, the nature of the Bi-Cu chemical bonding can still play a significant role in the brittle fracture of the grain boundary.

We have recently developed a phenomenological theory to separate the atomic size effect from the chemical contribution for substitutional impurities at the grain boundary in metals.[2] In our theory, the change of the work of grain boundary separation, $\Delta W_{sep}$, upon impurity segregation can be expressed as

(1) $\Delta W_{sep} = \frac{1}{3}(E_C^A - E_C^M + \Delta E_{sol}^A + \Delta E_{stru}^A) + \Delta E_V^A$,

where $E_C^A$ and $E_C^M$ are the cohesive energy of elemental crystal *A* (impurity) and *M* (matrix), $\Delta E_{sol}^A$ denotes the heat of formation of the alloy *A* in *M*, $\Delta E_{stru}^A$ represents the total energy difference of elemental crystal *A* between its ground state structure (rhombohedral for bismuth) and that of the host (face-centered cubic for Cu), and $\Delta E_V^A$ is the atomic size effect which, in the sphere and hole model,[4] takes the form

(2) $\Delta E_V^A = E_V^A - E_V^M = \frac{2K_A G_M (V^A - V^M - V^{GB})^2}{3K_A V^M + 4G_M V^A} - \frac{8K_A G_M (V^{GB})^2}{27K_A V^M + 36G_M V^A}$,

in which $K_A$ and $V^A$ are the bulk modulus and atomic volume of crystal *A*, $G_M$ and $V^M$ are the shear modulus and atomic volume of crystal *M*, and $V^{GB}$ is the free volume induced by grain boundary expansion available for *A* to accommodate itself. Highly precise first-principles all-electron full-potential linearized augmented plane wave method calculations on the effect of Co, Mo, Ru, Pd, W, and Re in Fe Σ3 (111) grain boundary[2,5] demonstrated that our phenomenological model works very well, an indication of the soundness of such a separation of atomic size and chemical bonding effects.

According to our model, only when the summation of the contributions in the parenthesis in Eq. 1 is smaller than zero, i.e., when the chemical effect is not embrittling, can we view the embrittling effect of an impurity as merely an atomic size effect. For the segregation geometry discussed by Schweinfest *et al.*, the corresponding Bi concentration in bulk Cu can be taken as 3/19.[6] The heat of formation $\Delta E_{sol}^A$ of such an alloy is 0.18 eV[7] and the structure term $\Delta E_{stru}^A$ is 0.14 eV. Hence, the total chemical contribution to the embrittling effect is 0.56 eV. Schweinfest *et al.* did not report the calculated volume expansion of Cu Σ19 (331). A reasonable estimate of the free volume available for Bi at this grain boundary is 15-30% of the atomic volume of bulk Cu. Then, from Eq.(2), the atomic size effect will be 0.95-1.12 eV, and the total embrittling effect of Bi is 1.51-1.68 eV – in very good agreement with their first-principles result, 1.57 eV (1.77 $Jm^{-2}$). Similarly, our model predicts a chemical contribution of 0.57 eV and an atomic size effect of 0.78-0.98 eV for an impurity of lead. The total embrittling potency, 1.35-1.55 eV, is also in very good agreement with their first-principles result 1.33 eV (1.51 $Jm^{-2}$).

Clearly, the numerical result of Schweinfest *et al.*'s work is strong support of our phenomenological theory. However, since they ignored the chemical contribution to the binding energy change of Bi, which contributes 35% of the calculated embrittlement potency, their attribution of its total embrittling effect only to its large atomic size is imprecise.

[1] Schweinfest, R., Paxton, A. T. & Finnis, M. W. *Nature* **432**, 1008-1011 (2004).

[2] Geng, W. T., Freeman, A. J. & Olson, G. B. Phys. Rev. B **63**, 165415 (2001).

[3] Rice, J. R. & Wang, J. -S. Mater. Sci. Eng. A **107**, 23-40 (1989).

[4] Friedel, J. Adv. Phys. **3**, 446 (1954).

[5] Geng, W. T., Freeman, A. J., Wu, R. & Olson, G. B. Phys. Rev. B **62**, 6208-6214 (2000).

[6] Sutton, A. P. & Vitek, V. *Acta Metall.* **30**, 2011-2033 (1982).

[7] Hultgren, R. *et al. Selected Values of the Thermodynamic Properties of Binary Alloys* (American Society for Metals, 1973)